\documentclass[12pt]{article}
\usepackage{amsmath,epsfig,graphicx,amssymb,mathrsfs}
\usepackage[margin=0.7in]{geometry}
\date{}
\begin{document}

\title{ A Cosmology Inspired Unified Theory of Gravity and Electromagnetism: Classical and Quantum Aspects}
\author{Partha Ghose\footnote{partha.ghose@gmail.com} \\
The National Academy of Sciences, India,\\ 5 Lajpatrai Road, Allahabad 211002, India}
\maketitle
\begin{abstract}
Milne cosmology has recently been shown to be in broad agreement with most cosmological data while being free of the problematic notions of standard cosmology such as the dark sector. In this paper a broken symmetric unified theory of gravity and electromagnetism is introduced which has a Milne metric under a certain geometric condition. Strikingly, particles (dyons) emerge as topological charges in this theory provided the torsion vector $\Gamma_i$ is curl-less.  
\end{abstract}
\section{Introduction}
Einstein's famous equation
\begin{equation}
R_{ik} - \frac{1}{2}g_{ik}R = \kappa T_{ik}\label{E}
\end{equation}
of General Relativity has been extremely successful in explaining and predicting various weak field phenomena such as the precession of the perihelion of Mercury, the bending of light by stars, the Shapiro delay time and the frame-dragging precession of gyroscopes measured by the Gravity Probe B experiment \cite{will}. In the strong field sector the observation of gravitational waves originating in the merger of binary black holes \cite{gwaves} has also come as a reassurance of the general correctness of the theory. 

The successful weak field predictions are all solutions of the field equations $R_{ik} = 0$, i.e. with $T_{ik} =0$ `outside' a spherically symmetric body having mass and angular momentum, such as the Schwarzschild and Kerr metrics. Hence, $R_{ik} = 0$ are not necessarily `vacuum equations' in the sense of a completely empty universe. On the other hand, attempts to use solutions of eqn (\ref{E}) with $T_{ik} \neq 0$ in FLRW cosmology have led to many intractable problems such as the hypothetical non-baryonic dark matter \cite{dm}, dark energy and the cosmological constant problem \cite{peebles}, the horizon problem \cite{hor} and the flatness problem \cite{flat} which show no signs of going away.

In this context the fact that the Milne model \cite{milne, milne2}, which has no horizon problem, happens to be in broad agreement with most cosmological data without requiring the dark sector of the concordance $\Lambda$CDM cosmology \cite{vish, vish2, mac} comes as a surprise and points to a possible alternative paradigm. This model is, however, based on Special Relativity but is {\em formally} identical with the so-called `vacuum' FLRW cosmology. It is therefore worth exploring if there exist some conditions under which it can be derived from a generalization of GR (a unified theory) without implying an empty universe.

Einstein himself was very unhappy with the role that the stress-energy tensor $T_{ik}$ played in GR. He had repeatedly emphasized that it was only a phenomenological representation of
matter, to be regarded with caution. In 1936 he wrote \cite{Ein}:
\begin{quote}
``[General Relativity] is sufficient--as far as we know--for the representation of the observed facts of celestial mechanics. But it is similar to a building, one wing of which is made of fine marble (left part of the equation), but the other wing of which is built of low-grade wood (right side of the equation). The phenomenological representation of matter is, in fact, only a crude substitute for a representation which would do justice to all known properties of matter.
\end{quote}
Also, in a letter to Michele Besso he wrote \cite{einarch}: 
\begin{quote}
``But it is questionable whether the equation $R_{ik} - \frac{1}{2} g_{ik} R = T_{ik}$ has any reality left within it in the face of quanta. I vigorously doubt it. In contrast, the left-hand side of the equation surely contains a deeper truth. If the equation $R_{ik} = 0$ really determines the behavior of the singularities, then a law describing this behavior would be justified far more deeply than the aforementioned equation, which is not unified and only phenomenologically justified.''
\end{quote} 
These quotes show that the stress-energy tensor was unsatisfactory to Einstein for two reasons. First, it is not geometrical in nature like the left side of eqn (\ref{E}) and hence not unified with it, and second, it does not reflect the quantum nature of matter and radiation. It is merely phenomenological, a placeholder for a more satisfactory theory of matter. This is why later on Einstein preferred to work with the equation $R_{ik} = 0$ in which matter appears as singularities and follow geodesics, though even this was a placeholder for a more satisfactory future theory of matter \cite{leh, Ein2}. 

It was therefore natural for him to try and construct a unified geometrical theory of all fields with the hope that the quantum features would emerge as consequences of the nonlinearity of the theory. The only known long range interactions being electromagnetic and gravity, he sought to bring them under one umbrella. Now, in General Relativity the number of independent variables is ten (the ten components of the metric tensor $g_{\mu\nu}$). Hence, in order to incorporate electromagnetism into a unified theory, one needed additional variables. There were many options for this. After Weyl's and Kaluza's attempts at unification, it was Eddington \cite{ed} who first proposed to replace the metric as a fundamental concept by a non-symmetric affine connection $\Gamma$ which could then be split into a symmetric and an anti-symmetric part. However, Einstein \cite{ein0}, supported by Schr\"{o}dinger \cite{sch}, extended the idea to include also a non-symmetric metric $g$. Just as passing beyond Euclidean geometry gravitation makes its appearance, so going beyond Riemannian geometry electromagnetism appears {\em naturally} as the anti-symmetric part of the metric {\em without requiring any higher dimensional space}. 

A major problem with such unified theories is that the symmetry between gravity and electromagnetism is actually badly broken in the universe, the electromagnetic interaction being enormously ($\sim 10^{36}$) stronger than gravity, but this problem is not usually addressed. The purpose of this paper is to consider a broken-symmetric unified theory with the aim of exploring if, under certain conditions, it leads to the Milne metric. To lead up to it, a minimal unified theory will be presented in the next section to set the scene for the broken-symmetric theory.

\section{A Minimal Unified Theory of Gravity and Electromagnetism}
Let $U_4$ be a smooth manifold with signature $(-,+,+,+)$ and endowed with a non-symmetric affine connection $\Gamma$ and a non-symmetric metric $g$.
Let
\begin{equation}
R_{ik} = \Gamma^\alpha_{ik,\,\alpha} - \Gamma^\alpha_{i\alpha,\,k} + \Gamma^\xi_{ik}\Gamma^\lambda_{\xi\lambda} - \Gamma^\xi_{i\lambda}\Gamma^\lambda_{\xi k} \label{R}
\end{equation}
be the non-symmetric curvature tensor. Let us also define
\begin{eqnarray}
\Gamma^\lambda_{(ik)} &=& \frac{1}{2}\left(\Gamma^\lambda_{ik}  + \Gamma^\lambda_{ki}\right),\\
\Gamma^{\lambda}_{[ik]} &=& \frac{1}{2}\left(\Gamma^\lambda_{ik}  - \Gamma^\lambda_{ki}\right),\\
\Gamma_l &=& \frac{1}{2}(\Gamma^\lambda_{l\lambda} - \Gamma^\lambda_{\lambda\ l}).
\end{eqnarray} 
$\Gamma^\lambda_{[ik]}$ is called the Cartan torsion tensor. Following Einstein \cite{ein}, let us put $\Gamma_l = 0$ since it is not determined by any equation in the theory. Similarly, let
\begin{eqnarray}
\bar{g}^{(ik)} &=& \frac{1}{2}\sqrt{-g}\left(g^{ik} + g^{ki}\right),\\
\bar{g}^{[ik]} &=& \frac{1}{2}\sqrt{-g}\left(g^{ik} - g^{ki}\right). 
\end{eqnarray}
To restrict the number of possible covariant terms in a non-symmetric theory, Einstein and Kaufman \cite{ein} imposed {\em transposition invariance} and {\em$\Lambda$-transformation invariance} on the theory.
Let $\tilde{\Gamma}^\lambda_{ik} = \Gamma^\lambda_{ki}$ and $\tilde{g}_{ik} = g_{ki}$. Then terms that are invariant under the simultaneous replacements of $\Gamma^\lambda_{ik}$ and $g_{ik}$ by $\tilde{\Gamma}^\lambda_{ik}$ and $\tilde{g}_{ik}$ are called transposition invariant. For example, the tensor $R_{ik}$ (\ref{R}) is not transposition invariant because it is transposed to
\begin{equation}
\tilde{R}_{ik} = \Gamma^\alpha_{ki,\,\alpha} - \Gamma^\alpha_{\alpha i,\,k} + \Gamma^\xi_{ki}\Gamma^\lambda_{\xi\lambda} - \Gamma^\xi_{\lambda i}\Gamma^\lambda_{k\xi}. \label{R2}
\end{equation}
Next, define the transformations
\begin{eqnarray}
\Gamma^{i\prime}_{kl} &=& \Gamma^{i}_{kl} + \delta^i_k \lambda_{,\,l},\nonumber\\
g^{ik\prime} &=& g^{ik}\label{proj},
\end{eqnarray}  
where $\lambda$ is an arbitrary function of the coordinates. Then  $R_{ik}$ (eqn \ref{R}) is $\lambda$-transformation invariant (or projective invariant). What this means is that a theory characterized by $R_{ik}$ cannot determine the $\Gamma$-field completely but only up to an arbitrary function $\lambda$. Hence, in such a theory, $\Gamma$ and $\Gamma^\prime$ represent the same field. Further, this {\em $\lambda$-transformation} produces a non-symmetric $\Gamma^\prime$ from a $\Gamma$ that is symmetric or anti-symmetric in the lower indices. Hence, the symmetry condition for $\Gamma$ loses objective significance. This sets the ground for a genuine unification of gravity and electromagnetism, the former determined by the symmetric part and the latter by the antisymmetric part of the action.

Let us write the simplest transposition invariant and $\lambda$-transformation invariant Lagrangian \cite{bose} 
\begin{eqnarray}
{\cal{L}} &=& \frac{1}{2}\sqrt{-g}\left(g^{ik}R_{ik} + \tilde{g}^{ik}\tilde{R}_{ik}\right)\label{L},
\end{eqnarray}
which can be expressed as (vide Appendix)
\begin{eqnarray}
{\cal{L}} &=& \bar{g}^{(ik)}\left(R_{(ik)} - \Gamma^\lambda_{[i\xi]}\Gamma^\xi_{[\lambda k]}\right) + \bar{g}^{[ik]}\Gamma^\lambda_{[ik];\,\lambda}\\
&:=& \bar{g}^{(ik)}R^\prime_{(ik)} + \bar{g}^{[ik]}\Gamma^\lambda_{[ik];\,\lambda}. 
\end{eqnarray}
Variation of the action $\int {\cal{L}} d^4 x$ holding $\bar{g}^{(ik)}$ and $\bar{g}^{[ik]}$ constant at once implies
\begin{eqnarray}
R^\prime_{(ik)} &=& R_{(ik)}- \Gamma^\lambda_{[i\xi]}\Gamma^\xi_{[\lambda k]} = 0,\label{1}\\
\Gamma^\lambda_{[ik];\,\lambda} &=& 0\label{2}.
\end{eqnarray}
It can also be shown (vide Appendix) using the variational principle that the equation connecting $g$ and $\Gamma$ in this non-symmetric theory is
\begin{eqnarray}
\bar{g}^{ik}_{\,\,\,\,\,,\lambda} &+& \bar{g}^{i\alpha}\Gamma^{\prime\, k}_{\lambda\alpha} +  \bar{g}^{\alpha k}\Gamma^{\prime\, i}_{\alpha\lambda} - \bar{g}^{ik}\Gamma^\alpha_{(\lambda\alpha)} = 0,\label{3}
\end{eqnarray}
where
\begin{eqnarray}
\Gamma^{\prime\, k}_{\lambda\alpha} &=& \Gamma^{k}_{(\lambda\alpha)} + \Gamma^{k}_{[\lambda\alpha]},\nonumber\\
\Gamma^{\prime\, i}_{\alpha\lambda} &=& \Gamma^{i}_{(\alpha\lambda)} + \Gamma^{i}_{[\alpha\lambda]},\nonumber
\end{eqnarray}
and further that
\begin{equation}
\bar{g}^{[i\alpha]}_{\,\,\,\,\,\,\,\,\,\,,\,\alpha} = 0.\label{4}
\end{equation}
The last equation can be interpreted as Maxwell's equations for electrodynamics by identifying $\bar{g}^{[ik]}$ with the dual electromagnetic field $\tilde{F}^{ik}$. The electric current is given by
\begin{eqnarray}
j^i &=& \frac{1}{3!}\zeta\epsilon^{i\nu\lambda\rho} \left(\bar{g}_{[\nu\lambda],\,\rho} + \bar{g}_{[\lambda\rho],\,\nu} + \bar{g}_{[\rho\nu],\,\lambda}\right)\nonumber\\
&=&\frac{1}{3!}\epsilon^{i\nu\lambda\rho}\left(\tilde{F}_{\nu\lambda,\,\rho} + \tilde{F}_{\lambda\rho,\,\nu} + \tilde{F}_{\rho\nu,\,\lambda}\right)\nonumber\\
&=& F^{i\nu}_{\,\,\,\,,\,\nu} = 0,\label{current}
\end{eqnarray}
where $\zeta$ is a suitable dimensional constant and the current vanishes because of the Bianchi identity in the first line. Hence, this theory describes free electromagnetic fields.

The equation set (\ref{1}), (\ref{2}), (\ref{3}), (\ref{4}) are the fundamental equations of the theory. 

Notice that that the Ricci tensor $R^\prime_{(ik)}$ in the theory, which is flat, has an additional term compared to the GR Ricci tensor $R_{(ik)}$. Clearly, the additional curvature has its origin in the torsion in the manifold $U_4$.  
\section{Broken Symmetric Unified Theory}
The theory outlined so far unifies gravity and electromagnetism fully, but in nature, as we have seen, these two interactions are distinguished by their widely different strengths, signalling a broken symmetry. One way to break the symmetry explicitly is to admit terms into the Lagrangian that are not $\lambda$-transformation or projective invariant. To have such a theory one needs to relax the Einstein condition $\Gamma_l = 0$. A general Lagrangian of this kind is \cite{bose}
\begin{eqnarray}
{\cal{L}} &=& \sqrt{-g}\left[g^{ik}R_{ik} + \tilde{g}^{ik}\tilde{R}_{ik}\right] + a\bar{g}^{(ik)}\Gamma_i\Gamma_k + b \bar{g}^{[ik]}\left(\Gamma_{i,\,k} - \Gamma_{k,\,i}\right) \label{L2}
\end{eqnarray}
where $a$ and $b$ are two arbitrary dimensionless parameters that are not fixed by any symmetry. Projective invariance requires $a$ to vanish but not $b$. Then eqns (\ref{1}) and (\ref{2}) are modified to
\begin{eqnarray}
{\cal{R}}_{ik} = R_{(ik)}- Q^\lambda_{i\xi}Q^\xi_{\lambda k} + x \Gamma_i\Gamma_k\nonumber\\ =
R_{(ik)}- \Gamma^\lambda_{[i\xi]}\Gamma^\xi_{[\lambda k]} + a \Gamma_i\Gamma_k &=& 0,\label{mod1}\\
Q^\lambda_{[ik];\,\lambda} - y\left(\Gamma_{i,\,k} - \Gamma_{k,\,i}\right) &=& 0,
\end{eqnarray}
where
\begin{eqnarray}
Q^\lambda_{ik} &=& \Gamma^{\,\,\,\,\,\,\,\,\,\,\lambda}_{[ik]} + \frac{1}{3}\delta^\lambda_i \Gamma_k - \frac{1}{3}\delta^\lambda_k \Gamma_i,\\
Q^\lambda_{ik;\,\lambda} &=& Q^\lambda_{ik,\,\lambda} - Q^\lambda_{i\xi}\Gamma^\xi_{(\lambda k)} - Q^\lambda_{\xi k}\Gamma^\xi_{(i\lambda)} + Q^\xi_{ik}\Gamma^\lambda_{(\xi\lambda)},
\end{eqnarray}
and $x = a + \frac{1}{3}$, $y = \frac{1}{6} - b$. Hence, the new Ricci tensor ${\cal{R}}_{ik}$ is also flat though the universe has other fields than gravity, and is hence not empty.

Eqn (\ref{mod1}) will lead to corrections to the Schwarzschild and Kerr metrics that are analogous to the Reissner-Nordstr\"{o}m and Kerr-Newman metrics.

The variational principle is a little more complex because $Q^\lambda_{i\lambda} = 0$ and all the 24 components of $Q^\lambda_{ik}$ are not independent, and consequently one has to use a Lagrange multiplier $k^i$ \cite{bose}. The upshot is that the equations connecting the $g$'s and $\Gamma$'s are of the form
\begin{eqnarray}
\bar{g}^{ik}_{\,\,\,\,\,,\lambda} + \bar{g}^{i\alpha}\Gamma^{\prime\prime\, k}_{\lambda\alpha} +  \bar{g}^{\alpha k}\Gamma^{\prime\prime\, i}_{\alpha\lambda} &=& 3 \bar{g}^{ik}\Phi_\lambda,\label{eqcon}\\
\Phi_\lambda = g_{[\lambda\beta]}k^\beta &=& -\frac{1}{3}\left(\frac{x}{y}\right) g_{[\lambda\beta]}g^{(\beta\alpha)}\Gamma_\alpha,
\end{eqnarray}
where
\begin{eqnarray}
\Gamma^{\prime\prime\, k}_{\lambda\alpha} &=& \Gamma^{k}_{(\lambda\alpha)} + Q^{k}_{\lambda\alpha} + \frac{1}{\sqrt{\vert g\vert}}(g_{\lambda\beta}k^\beta \delta^k_\alpha - g_{\beta\alpha}k^\beta \delta^k_\lambda).\nonumber\\
\end{eqnarray}
Equation (\ref{4}) is modified to
\begin{equation}
\bar{g}^{[i\alpha]}_{\,\,\,\,\,\,\,\,,\alpha} = 3k^i = - \frac{x}{y}\bar{g}^{(i\alpha)}\Gamma_\alpha := \theta \Gamma^i,\,\,y\neq 0.\label{curr}
\end{equation}
Thus, $\Gamma^i$ turns out to be the source of the dual electromagnetic field unless $x = 0$, i.e. $a = -\frac{1}{3}$ and $k^i = 0$. If $\Gamma_l = k_l = 0$, one gets back the minimally unified theory. By multiplying the equation by the dimensional parameter $\zeta$, we can write it in the form
\begin{equation}
\tilde{F}^{i\alpha}_{\,\,\,\,\,\,\,\,,\alpha} = j^i_m \label{jm}
\end{equation}
where $j^i_m = \zeta\theta\Gamma^i$ is the magnetic source current which is automatically conserved because $\tilde{F}^{i\alpha} = -\tilde{F}^{\alpha i}$.
Defining $F_{kl} = \epsilon_{kli\alpha}\tilde{F}^{i\alpha}$, we have
\begin{equation}
\partial^k F_{kl} = \partial^k\epsilon_{kli\alpha}\tilde{F}^{i\alpha} := j_l \label{j}
\end{equation}
where $j_l$ is the electric source current which is also automatically conserved. These two equations ((\ref{jm}), (\ref{j})) constitute the complete set of Maxwell equations in the presence of the two source currents:
\begin{eqnarray}
\vec{\nabla} \times \vec{B} - \frac{\partial \vec{E}}{\partial t} &=& \vec{j},\,\,\,\,\,\,\vec{\nabla}. \vec{E} = \rho_e,\label{ma1}\\
\vec{\nabla} \times \vec{E} + \frac{\partial \vec{B}}{\partial t} &=& - \vec{j}_m,\,\,\,\,\,\,\vec{\nabla}. \vec{B} = \rho_m.\label{ma2}
\end{eqnarray}
The two Bianchi identities that must be satisfied are
\begin{eqnarray}
F_{\mu\nu,\,\lambda} + F_{\nu\lambda,\,\mu} + F_{\lambda\mu,\,\nu} &=& 0, \label{B1}\\
\tilde{F}_{\mu\nu,\,\lambda} + \tilde{F}_{\nu\lambda,\,\mu} + \tilde{F}_{\lambda\mu,\,\nu} &=& 0. \label{B2} 
\end{eqnarray}
These identities are consistent with the inhomogeneous Maxwell equations provided one makes the following identifications:
\begin{eqnarray}
F^{0i} &=& -(E^i - E^{\prime\,i}),\,\,\,\,\,\, F^{ij} = - \epsilon^{ijk} (B_k - B_k^\prime),\\
\tilde{F}^{0i} &=& - (B^i - B^{\prime\,i}),\,\,\,\,\,\,\tilde{F}^{ij} = \epsilon^{ijk}(E_k - E^\prime_k), 
\end{eqnarray}
where $(\vec{E}^\prime, \vec{B}^\prime)$ are auxiliary fields that satisfy the conditions
\begin{eqnarray}
\vec{\nabla}.\vec{E}^{\prime} &=& \rho_e,\,\,\,\,\vec{\nabla}\times\vec{E}^{\prime} = -\vec{j}_m,\\
\vec{\nabla}.\vec{B}^{\prime} &=& \rho_m,\,\,\,\,\vec{\nabla}\times\vec{B}^{\prime} = \vec{j},\\
\frac{\partial \vec{E}^{\prime}}{\partial t} &=& \frac{\partial \vec{B}^{\prime}}{\partial t} = 0. 
\end{eqnarray}
One can now define potentials $A^\mu = (\phi, \vec{A}), \tilde{A}^\mu = (\tilde{\phi}, \vec{\tilde{A}})$ through the relations
\begin{eqnarray}
\vec{E} &=& -\frac{\partial \vec{A}}{\partial t} - \vec{\nabla}\phi,\\
\vec{B} &=& \vec{\nabla}\times \vec{A},\\
\vec{E}^{\prime} &=& \vec{\nabla}\times \vec{\tilde{A}} + \vec{\nabla}\phi, \\
\vec{B}^{\prime} &=& \vec{\nabla}\times \vec{A} + \vec{\nabla}\tilde{\phi}
\end{eqnarray}
in the Lorentz gauge $\vec{\nabla}.\vec{A} = \vec{\nabla}.\vec{\tilde{A}} = 0$ and with $\nabla^2\tilde{\phi} = \rho_m,\,\,\nabla^2\phi = \rho_e, \nabla^2\vec{A} = -\vec{j}, \nabla^2\vec{\tilde{A}} = \vec{j}_m$.
Hence, {\em the theory allows magnetic charges and currents without Dirac strings} \cite{dir}.

An interesting feature of the presence of $\Gamma_i$ is that it makes electrodynamics invariant under continuous transformations \cite{heav, lar} 
\begin{eqnarray}
\vec{E}\rightarrow\vec{E}^\prime &=& \vec{E}{\rm cos}\theta - \vec{B}{\rm sin}\theta,\\
\vec{B}\rightarrow \vec{B}^\prime &=& \vec{E}{\rm sin}\theta + \vec{B}{\rm cos}\theta,
\end{eqnarray}
where $0\leq\theta\leq \pi/2$. Hence,
\begin{eqnarray}
\vec{j}^\prime &=& \vec{j}{\rm cos}\theta - \vec{j}_m{\rm sin}\theta,\\
\vec{j}_m^\prime &=& \vec{j}{\rm sin}\theta + \vec{j}_m{\rm cos}\theta,\\
\rho_e^\prime &=& \rho_e{\rm cos}\theta - \rho_m{\rm sin}\theta,\\
\rho_m^\prime &=& \rho_e{\rm sin}\theta + \rho_m{\rm cos}\theta.
\end{eqnarray}
For $\theta = \pi/2$ one has
$\vec{E} \rightarrow -\vec{B}, \vec{B} \rightarrow \vec{E},
(\rho_e, \vec{j}) \rightarrow (-\rho_m, -\vec{j}_m),
(\rho_m, \vec{j}_m) \rightarrow (\rho_e, \vec{j})$. This shows that there is complete equivalence and continuous freedom in the choice of
electric and magnetic quantities.

\subsection{Quantization}

Note that $(\rho_m, \rho_e)$ are time components of the corresponding 4-currents which are determined in terms of continuous fields (vide eqns (\ref{current}) and (\ref{curr})). Also, note that no condition has been imposed so far on $\Gamma_i$. Instead of imposing the Einstein condition $\Gamma_i = 0$, if one imposes the weaker condition
\begin{equation}
\Gamma_{i,\,k} - \Gamma_{k,\,i} = 0,\label{cond}
\end{equation}
then like ${\cal{R}}_{ik} = 0$, $Q^\lambda_{[ik];\,\lambda} = 0$, and
one immediately gets a very interesting result, namely that $\Gamma_i$, and hence the magnetic source current $j_{m\,i} = \zeta \Gamma_i$, is an {\em irrotational} or curl-less axial vector. Let $S = \mathbb{R}^3\backslash \left\{(0,0,z\leq 0)|z \in \mathbb{R}\right\}$ be the usual 3-dimensional space with the negative $z$-axis, along which $\vec{\Gamma} \neq 0$, removed. Then the curl-less vector $\vec{\Gamma} = -\vec{\nabla}\tilde{\Phi},\,\nabla^2 \tilde{\Phi} = 0$ has vortex solutions $\vec{\Gamma} = \vec{e}_\phi/r$ where $\vec{e}_\phi$ is a unit vector, and the integral over a unit counterclockwise circular path $C$ in the $xy$ plane enclosing the origin is 
\begin{equation}
\frac{1}{2\pi}\oint_C \vec{\Gamma}.\vec{e}_\phi\, d\phi =  n,\,\,n\in \mathbb{Z},
\end{equation}
where $n$ is a winding number which can be interpreted as the number of magnetic charges enclosed by the unit circle. For $n = 1$ one has a single magnetic charge, i.e. a magnetic monopole carrying some charge $g$. It is a straightforward consequence of condition (\ref{cond}). Because of the Larmor-Heaviside symmetry, there is also an electric monopole, i.e. a particle carrying electrical charge $e$. Thus, particles emerge as topological charges. The product $eg$ has the dimension of action, and the fundamental unit of action in nature being the Planck constant $h$, it follows that $eg/h =$ constant, which is essentially the topological basis of Dirac quantization.

\section{Cosmological Implications}
We have seen that in the broken symmetric unified theory the Ricci tensor ${\cal{R}}_{ik}$ is flat. This will have non-trivial implications for FLRW cosmology whose metric is
\begin{equation}
ds^2 = dt^2 - a^2(t)\left[\frac{dr^2}{1 - kr^2} + r^2 d\theta^2 + r^2 {\rm sin}^2\theta d\phi^2 \right].
\end{equation}
In GR the spatial curvature $k$ is related to the Ricci scalar $R = g^{(ik)}R_{ik}$ by the relation $k = -a^2R/6$. In the broken symmetric unified theory this relation is modified to the form
\begin{equation}
k = - \frac{a^2}{6}(R - \Gamma_T + a\Gamma)
\end{equation}
where 
\begin{eqnarray}
\Gamma_T &=& g^{(ik)}\Gamma^\lambda_{[i\xi]}\Gamma^\xi_{[\lambda k]},\,\,\Gamma = g^{(ik)}\Gamma_i\Gamma_k.
\end{eqnarray}
Hence $k = -1$ provided 
\begin{equation}
R -\Gamma_T + a\Gamma = 6/a^2.\label{X}
\end{equation}
This is therefore the required condition to have the Milne metric in a broken symmetric unified theory which reduces to GR in the absence of torsion. Notice that the additional torsional terms in ${\cal{R}}_{ik}$ (\ref{mod1}) are due to the presence of electromagnetic fields and charged particles in the universe which is therefore not empty. In principle it is possible to express them in terms of electromagnetic fields and currents by using solutions of the equation (\ref{eqcon}) which give the connections in terms of the metrics, but they turn out to be very complex \cite{bose2, ton, hlav}.
\section{Concluding Remarks}
A remarkable feature of the broken symmetric unified theory with $\Gamma_i \neq 0$ is the occurrence of particles as topological charges carrying electric and magnetic charge provided $\Gamma_i$ is curl-less. Thus, Einstein's dream of deriving particles from fields is at least partially realized.

Furthermore, the theory implies Milne cosmology if condition (\ref{X}) holds. Milne cosmology is consistent with with most cosmological data without requiring the dark sector of the concordance $\Lambda$CDM cosmology.

Because of the presence of the magnetic current $\Gamma_l$ there is also a natural classical explanation of magnetism in the universe. Magnetism is ubiquitous in the universe, and primordial magnetic fields are specially important for probing the physics of
the early universe \cite{kanda}.
\section{Acknowledgement}
The author thanks the National Academy of Sciences, India for a grant.
\section{Appendix}
Straightforward algebra gives
\begin{eqnarray}
{\cal{L}} &=& \frac{1}{2}\sqrt{-g}\left(g^{ik}R_{ik} + \tilde{g}^{ik}\tilde{R}_{ik}\right)\nonumber\\ 
&=& \left[\bar{g}^{(ik)}(R_{ik} + \tilde{R}_{ki}) + \bar{g}^{[ik]}(R_{ik} - \tilde{R}_{ki})\right]\nonumber\\
&=& \bar{g}^{(ik)}\left[R_{ik}  - \Gamma^{\,\,\,\,\,\,\,\,\,\,\lambda}_{[i\xi]}\Gamma^{\,\,\,\,\,\,\,\,\,\,\xi}_{[\lambda k]}\right]\nonumber\\ &+& \bar{g}^{[ik]}\left(\Gamma^\lambda_{[ik],\,\lambda} -\Gamma^\lambda_{[i\lambda],\,k} +\Gamma^\xi_{ik}\Gamma^\lambda_{\xi\lambda} - \Gamma^\xi_{i\lambda}\Gamma^\lambda_{\xi k} - \Gamma^\xi_{ki}\Gamma^\lambda_{\lambda\xi} + \Gamma^\xi_{\lambda i}\Gamma^\lambda_{k\xi}\right)\nonumber\\
&=&  \bar{g}^{(ik)}\left[R_{ik} - \Gamma^{\,\,\,\,\,\,\,\,\,\,\lambda}_{[i\xi]}\Gamma^{\,\,\,\,\,\,\,\,\,\,\xi}_{[\lambda k]}\right] + \bar{g}^{[ik]}\left[\Gamma^\lambda_{ik;\,\lambda})\right]
\end{eqnarray}
Following \cite{bose} but using $\Gamma_\mu = 0$, let
\begin{equation}
{\cal{L}} = H + \frac{d X^\lambda}{dx^\lambda}\label{H}
\end{equation} 
with
\begin{eqnarray}
X^\lambda &=& \bar{g}^{(ik)}\Gamma_{(ik)}^\lambda - \bar{g}^{(i\lambda)}\Gamma^k_{(ik)} + \bar{g}^{[ik]}\Gamma^\lambda_{[ik]}\nonumber\\
H &=& - \bar{g}^{(ik)}_{\,\,\,\,\,\,,\lambda}\Gamma^\lambda_{(ik)} + \bar{g}^{(i\lambda)}_{\,\,\,\,\,,\lambda}\Gamma^k_{{(ik)}} + \bar{g}^{(ik)}\left(\Gamma^\xi_{(ik)}\Gamma^\lambda_{(\xi\lambda)} - \Gamma^\xi_{(i\lambda)}\Gamma^\lambda_{(\xi k)} - \Gamma^\lambda_{[i\xi]}\Gamma^\xi_{[\lambda k]}\right)\nonumber\\ &-& \bar{g}^{[ik]}_{\,\,\,\,\,\,,\lambda}\Gamma^\lambda_{[ik]} + \bar{g}^{[ik]}\left[- \Gamma^\lambda_{[i\xi]}\Gamma^\xi_{(\lambda k)} - \Gamma^\lambda_{[\xi k]}\Gamma^\xi_{(i\lambda)} + \Gamma^\xi_{[ik]}\Gamma^\lambda_{(\xi\lambda)}\right] \nonumber
\end{eqnarray}
Thus, $H$ is free of the partial derivatives of $\Gamma^\lambda_{(ik)}$ and $\Gamma^\lambda_{[ik]}$, and the four-divergence term in the action integral is equal to a surface integral at infinity on which all arbitrary variations are taken to vanish. 

Variations of $H$ w.r.t $\Gamma^\lambda_{(ik)}$ and $\Gamma^\lambda_{[ik]}$ give
\begin{eqnarray}
\bar{g}^{(ik)}_{\,\,\,\,\,,\lambda} + \bar{g}^{(i\alpha)}\Gamma^k_{(\lambda\alpha)} + \bar{g}^{(\alpha k)}\Gamma^i_{(\alpha\lambda)} - \bar{g}^{(ik)}\Gamma^\alpha_{(\lambda\alpha)}  = -[\bar{g}^{[i\alpha]}\Gamma^{\,\,\,\,\,\,k}_{[\lambda\alpha]} + \bar{g}^{[\alpha k]}\Gamma^{\,\,\,\,i}_{[\alpha\,\,\,\,\lambda]}]\label{1a} \\
\bar{g}^{[ik]}_{\,\,\,\,\,,\lambda} + \bar{g}^{[i\alpha]}\Gamma^{k}_{(\lambda\alpha)} + \bar{g}^{[\alpha k]}\Gamma^{i}_{(\alpha\lambda)} - \bar{g}^{[ik]}\Gamma^\alpha_{(\lambda\alpha)} = - [\bar{g}^{(i\alpha)}\Gamma^{\,\,\,\,\,\,k}_{[\lambda\alpha]} + \bar{g}^{(\alpha k)}\Gamma^{\,\,\,\,i}_{[\alpha\,\,\,\,\lambda]}]\label{2a}
\end{eqnarray}  
Adding (\ref{1a}) and (\ref{2a}), we get
\begin{eqnarray}
\bar{g}^{ik}_{\,\,\,\,,\lambda} &+& \bar{g}^{i\alpha}\left(\Gamma^k_{(\lambda\alpha)} + \Gamma^{k}_{[\lambda\alpha]}\right) +  \bar{g}^{\alpha k}\left(\Gamma^i_{(\alpha\lambda)} + \Gamma^{i}_{[\alpha\lambda]}\right) - \bar{g}^{ik}\Gamma^\alpha_{(\lambda\alpha)} = 0.
\end{eqnarray} 
This can be written as
\begin{eqnarray}
\bar{g}^{ik}_{\,\,\,\,\,,\lambda} &+& \bar{g}^{i\alpha}\Gamma^{\prime\, k}_{\lambda\alpha} +  \bar{g}^{\alpha k}\Gamma^{\prime\, i}_{\alpha\lambda} - \bar{g}^{ik}\Gamma^\alpha_{(\lambda\alpha)} = 0,\label{ggamma}\\
\Gamma^{\prime\, k}_{\lambda\alpha} &=& \Gamma^{k}_{(\lambda\alpha)} + \Gamma^{k}_{[\lambda\alpha]},\nonumber\\
\Gamma^{\prime\, i}_{\alpha\lambda} &=& \Gamma^{i}_{(\alpha\lambda)} + \Gamma^{i}_{[\alpha\lambda]}.\nonumber
\end{eqnarray}
This is eqn (\ref{3}).

By contracting (\ref{ggamma}) once with respect to $(k, \lambda)$, then with respect to $(i, \lambda)$, and subtracting the equations term by term, one gets eqn (\ref{4}).

\end{document}